\documentclass[letter]{aa}
\usepackage{amsmath}
\usepackage{mathrsfs}
\usepackage{dblfloatfix}
\usepackage{color}
\usepackage{setspace}
\usepackage{graphicx}
\usepackage{fixltx2e}
\usepackage{float}
\usepackage{txfonts}
\usepackage{placeins}
\usepackage{array, makecell} 
\usepackage{gensymb}
\usepackage{booktabs}
\usepackage{url}
\usepackage{subfloat}
\usepackage[version=3]{mhchem}
\usepackage{threeparttable}
\usepackage{multirow}
\usepackage{CJK}
\usepackage{longtable}
\usepackage{pdflscape}
\usepackage[export]{adjustbox}
\usepackage{subfig}
\usepackage{array}
\usepackage{bm}
\usepackage[singlelinecheck=false, labelfont=bf]{caption}

\usepackage{natbib,twoopt}
\usepackage[hyphenbreaks]{breakurl}
\usepackage[breaklinks]{hyperref}      
\bibpunct{(}{)}{;}{a}{}{,}             
\definecolor{cobalt}{rgb}{0.06, 0.2, 0.65}
\hypersetup{
  colorlinks,
  citecolor=cobalt,
  linkcolor=cobalt,
  urlcolor=cobalt,
}
\makeatletter
  \newcommandtwoopt{\citeads}[3][][]{\href{http://adsabs.harvard.edu/abs/#3}%
    {\def\hyper@linkstart##1##2{}%
     \let\hyper@linkend\@empty\citealp[#1][#2]{#3}}}
  \newcommandtwoopt{\citepads}[3][][]{\href{http://adsabs.harvard.edu/abs/#3}%
    {\def\hyper@linkstart##1##2{}%
     \let\hyper@linkend\@empty\citep[#1][#2]{#3}}}
  \newcommandtwoopt{\citetads}[3][][]{\href{http://adsabs.harvard.edu/abs/#3}%
    {\def\hyper@linkstart##1##2{}%
     \let\hyper@linkend\@empty\citet[#1][#2]{#3}}}
  \newcommandtwoopt{\citeyearads}[3][][]%
    {\href{http://adsabs.harvard.edu/abs/#3}
    {\def\hyper@linkstart##1##2{}%
     \let\hyper@linkend\@empty\citeyear[#1][#2]{#3}}}
\makeatother

\newcommand{\myemail}{\email{\href{mailto:cyang@eso.org}{cyang@eso.org}}}

\newcommand{\kms}{{\hbox {\,km\,s$^{-1}$}}}
\newcommand{\mum}{{\hbox {\,$\mu$m}}}
\newcommand{\kkmspc}{{\hbox {\,K\,km\,s$^{-1}$\,pc$^{2}$}}} 
\newcommand{\lsun}{{\hbox {$L_\odot$}}}
\newcommand{\msun}{{\hbox {$M_\odot$}}}

\newcommand{\hto}{{\hbox {H\textsubscript{2}O}}}

\newcommand{\lco}{\hbox {$L_{\mathrm{CO}}$}}

\newcommand{\lir}{\hbox {$L_{\mathrm{IR}}$}}

\label{key}

\def\co#1#2{{\hbox {${\mathrm{CO}}(#1\text{--}#2)$}}}

\def\lco#1#2{\hbox {$L_{\mathrm{CO}(#1\text{--}#2)}$}}

\def\htot#1#2#3#4#5#6{\hbox {\hto(\t#1#2#3#4#5#6)}}
\def\t#1#2#3#4#5#6{{\hbox {$#1_{#2#3}\text{--}#4_{#5#6}$}}}

\def\lhtot#1#2#3#4#5#6{\hbox {$L_{\mathrm{H_2O}(#1_{#2#3}\text{--}#4_{#5#6})}$}}

\defcitealias{2019A&A...624A.138Y}{Y19}

\begin{document}
\begin{CJK*}{UTF8}{gbsn}

\title{First detection of the 448\,GHz
ortho-{\texorpdfstring{H\textsubscript{2}O}{H2O}} 
line at high redshift: probing the structure of a starburst nucleus 
at \textit{\textbf{z}}\,=\,3.63}


\author
{
  \href{https://orcid.org/0000-0002-8117-9991}{C. Yang\,(杨辰涛)}\inst{1}         \and 
  \href{https://orcid.org/0000-0001-5285-8517}{E. Gonz{\'a}lez-Alfonso}\inst{2}   \and
  \href{https://orcid.org/0000-0002-4721-3922}{A. Omont}\inst{3}                  \and 
  \href{https://orcid.org/0000-0002-4005-9619}{M. Pereira-Santaella}\inst{4}      \and 
  \href{https://orcid.org/0000-0001-6697-7808}{J. Fischer}\inst{5}                \and   
  \href{https://orcid.org/0000-0003-3201-0185}{A. Beelen}\inst{6}                 \and  
  \href{https://orcid.org/0000-0002-5540-6935}{R. Gavazzi}\inst{3}                
}

\institute{
European Southern Observatory, Alonso de C{\'o}rdova 3107, Casilla 19001, 
Vitacura, Santiago, Chile. \myemail 
\and
Universidad de Alcal{\'a}, Departamento de F\'{\i}sica y Matem{\'a}ticas, Campus 
Universitario, 28871 Alcal{\'a} de Henares, Madrid, Spain
\and
Institut d'Astrophysique de Paris, UMR7095 CNRS \& Sorbonne Universit\'e (UPMC), 
98 bis bd Arago, 75014 Paris, France. 
\and  
Centro de Astrobiolog\'{\i}a (CSIC-INTA), Ctra. de Ajalvir, Km 4, 28850, 
Torrej\'on de Ardoz, Madrid, Spain
\and
George Mason University, Department of Physics \& Astronomy, MS 3F3, 4400
  University Drive, Fairfax, VA 22030, USA
\and
Institut d'Astrophysique Spatiale, CNRS UMR 8617, Universit\'{e} Paris-Sud, 
Universit\'{e} Paris-Saclay, 91405 Orsay, France
}

\date {Received .../ Accepted ...}

\abstract
{Submillimeter rotational lines of \hto\ are a powerful probe in  
warm gas regions of the interstellar medium (ISM), tracing scales 
and structures ranging from kiloparsec disks to the most compact 
and dust-obscured regions of galactic nuclei. The ortho-\htot423330\ 
line at 448\,GHz, which was recently detected in a local luminous 
infrared galaxy \citep{2017A&A...601L...3P}, offers a unique 
constraint on the excitation conditions and ISM properties in 
deeply buried galaxy nuclei since the line requires high far-infrared 
optical depths to be excited. In this letter, we report the first 
high-redshift detection of the 448\,GHz \htot423330\ line using 
ALMA, in a strongly lensed submillimeter galaxy (SMG) at $z=3.63$. 
After correcting for magnification, the luminosity of the 448\,GHz 
\hto\ line is $\sim 10^{6}$\,\lsun. In combination with 
three other previously detected \hto\ lines, we build a model that 
``resolves'' the dusty ISM structure of the SMG, and find that it 
is composed of a $\sim$\,1\,kpc optically thin (optical depth at 
100\,$\mu$m $\tau_{100}$\,$\sim$\,0.3) disk component with 
dust temperature $T_{\rm dust}\approx50$\,K emitting a total 
infrared power of $5\times10^{12}$\,\lsun\ with surface density 
$\Sigma_\mathrm{IR}=4\times10^{11}$\,\lsun\,kpc$^{-2}$, and 
a very compact (0.1\,kpc) heavily dust-obscured 
($\tau_{100}$\,$\gtrsim$\,1) nuclear core with very warm dust 
(100\,K) and $\Sigma_\mathrm{IR}=8\times10^{12}$\,\lsun\,kpc$^{-2}$. 
The \hto\ abundance in the core component, 
$X_\mathrm{H_{2}O}$\,$\sim$\,$(0.3\text{--}5)\times 10^{-5}$, 
is at least one order of magnitude higher than in the 
disk component. The optically thick core has the characteristic 
properties of an Eddington-limited starburst, providing evidence 
that radiation pressure on dust is capable of supporting the 
ISM in buried nuclei at high redshifts. The multi-component 
ISM structure revealed by our models illustrates that dust 
and molecules such as \hto\ are present in regions characterized 
by highly differing conditions and scales, extending from 
the nucleus to more extended regions of SMGs.
}


\authorrunning{C. Yang et al.}

\titlerunning{Probing the structure of the starburst nuclei in $z \sim$\,3.6 SMG with \hto\ lines}

\maketitle


\section{Introduction}

Either in the gas phase in warm regions, or locked onto dust 
mantles in cold environments, H$_2$O is one of the most 
abundant molecules in the interstellar medium (ISM). In 
addition to probing a variety of physical processes such as 
shocks \citep{2010MNRAS.406.1745F}, radiative pumping 
\citep{2008ApJ...675..303G} and outflowing gas 
\citep{2016A&A...593A..43V}, it plays an essential role 
in the oxygen chemistry of the ISM 
\citep[e.g.,][]{2013ChRv..113.9043V}.
Recent observations of the rotational transitions of the \hto\ 
lines in the submillimeter (submm) bands show their ubiquity 
in infrared (IR) bright galaxies and reveal the tight 
relation between the submm \hto\ lines and dust emission
\citep{2013ApJ...771L..24Y}. Case studies of local IR-bright 
galaxies have demonstrated that far-IR pumping plays an 
important role in the excitation of the \hto\ lines 
\citep[e.g.,][]{2010A&A...518L..43G, 2012A&A...541A...4G, 2017ApJ...846....5L}. 
The \hto\ lines offer diagnostics of regions of 
warm gas, which are usually deeply buried in dust, 
probing different properties of the ISM than are probed 
by collisionally excited lines like CO. By modeling the 
\hto\ excitation, dust properties such as the dust temperature, 
far-IR optical depth, and IR luminosity, can be constrained, 
and can even be decomposed into multiple components that 
reveal the structure of the dust-obscured ISM
\citep[e.g.,][]{2015A&A...580A..52F}. 

H$_2$O lines are a particularly powerful diagnostic tool for 
studying dusty galaxies. At high redshifts, such galaxies 
were first discovered in the submm and later characterized 
as submm galaxies \citep[SMGs, e.g.,][or dusty star-forming galaxies, 
DSFGs, \citealt{2014PhR...541...45C}]{1997ApJ...490L...5S}. 
SMGs are undergoing massive star formation, sometimes reaching the 
``maximum starburst'' limit \citep[e.g.,][]{2014ApJ...784....9B}. 
The extremely intense star formation suggests that SMGs 
are in the critical phase of rapid stellar mass assembly.  
They are likely linked to the local massive spheroidal galaxies 
\citep[e.g.,][]{2014ApJ...782...68T}. However, the nature of 
SMGs remains debated \citep[e.g.,][]{2010MNRAS.404.1355D, 
2015Natur.525..496N}, in part due to lack of spatially resolved 
studies of their dusty ISM. Heretofore, most of the \hto\ 
studies at high redshifts have been based on low spatial 
resolution observations, which reveal the average properties 
of the ISM in SMGs 
\citep[e.g.,][]{2013A&A...551A.115O, 2016A&A...595A..80Y}.
Nevertheless, with ALMA, \cite{2018ApJ...863...56C} find 
evidence of significant radial variation of the ISM properties 
in SMGs and suggest caution when interpreting single band 
dust continuum data. Furthermore, by measuring the 
structure of the dusty ISM, one can assess the stellar 
mass assembly history of SMGs and build a link to galaxy 
populations of today \citep[e.g.,][]{2019ApJ...879...54L}. 
However, such observations require resolving the continuum 
emission in multiple bands that sample the peak of the dust 
spectral energy distribution (SED) at high frequencies.  
Moreover, accessing the most dust-obscured regions is
observationally challenging. The \hto\ lines thus provide 
an alternative approach to constrain the structure and 
properties of the dusty ISM in SMGs, owing to their tight 
link to the far-IR radiation field.

The ortho-\htot423330\ line ($E_\mathrm{u}=433$\,K) at 
448.001\,GHz was recently detected for the first time in 
space, with the Atacama Large Millimeter/submillimeter Array 
(ALMA) in ESO\,320-G030, an isolated IR-luminous barred 
spiral that is likely powered by a starburst, since, 
based on X-ray and mid-IR diagnostics, 
there is no evidence for an obscured active galactic 
nucleus (AGN) \citep{2017A&A...601L...3P}. In this source, 
\htot423330\ is excited by intense far-IR radiation, 
rather than being a maser line as predicted by collisional 
models \citep[e.g.,][]{1991ApJ...368..215N,2016MNRAS.456..374G}. 
The spatially resolved observations of the \htot423330\ line 
and dust continuum in ESO\,320-G030 indicate that the 
line arises from a highly obscured galactic 
nucleus. Therefore, this highly excited \hto\ line is 
an ideal probe of the deeply buried (optically thick in 
the far-IR), warm dense nuclear ISM of galaxies. 

In this letter, we report the first high-redshift detection 
of the ortho-\htot423330\ line in a strongly lensed SMG 
at $z=3.63$ in the SMG merging pair G09v1.97. It was 
first discovered with {\it Herschel} \citep{2013ApJ...779...25B} 
and was followed up with low spatial resolution 
observations of several \hto\ and CO lines 
\citep{2016A&A...595A..80Y, 2017A&A...608A.144Y}, 
and high-resolution observation of the \htot211202\ and \co65\ lines 
\citep[][\citetalias{2019A&A...624A.138Y} hereafter]{2019A&A...624A.138Y}.
G09v1.97 has a total molecular gas mass of 10\,$^{11}$\,\msun, 
and is composed of two gas-rich merging galaxies. 
The two galaxies, dubbed G09v1.97-R (the 
receding northern galaxy) and G09v1.97-B (the approaching 
southern galaxy), are separated by a projected distance 
of 1.3\,kpc ($\sim$\,0\farcs2) and have total intrinsic IR 
luminosities (8--1000\,\mum) of $6.3\times10^{12}$ and 
$4.0\times10^{12}$\,\lsun, respectively.
While both G09v1.97-R and G09v1.97-B are one order of 
magnitude more powerful than ESO\,320-G030 in IR luminosity,
like ESO\,320-G030, these galaxies are likely powered by 
star-formation, based on their similar line-to-IR luminosity
ratios of \lco65/\lir\ and \lhtot211202/\lir\  
(\citealt{2013ApJ...771L..24Y}, \citetalias{2019A&A...624A.138Y}).

We adopt a spatially flat $\Lambda$CDM cosmology 
with $H_{0}=67.8\,{\rm km\,s^{-1}\,Mpc^{-1}}$, 
$\Omega_\mathrm{M}=0.308$ \citep{2016A&A...594A..13P},    
and a Chabrier \citeyearpar{2003PASP..115..763C} 
initial mass function (IMF) throughout this work.

\section{Observations and data reduction}
\label{section:obs_redu}
 
The ALMA observations presented here are part of 
a dense gas line survey project (ADS/JAO.ALMA\#2018.1.00797.S, 
Yang\,et al., in prep.). The observations were carried 
out between December 2018 and January 2019. In this work, 
we only present data from the Band-3 spectral window 
covering the ortho-\htot423330\ line centered 
at 96.137\,GHz (observed frequency). Forty-three antennas 
of the 12-m array were used. The observations were 
performed under good weather conditions 
(PWV\,=\,2--5\,mm, phase RMS\,$<$\,9$^\circ$) 
with the C43-2 configuration, which provides 
baselines ranging from 15 to 200\,m. J0825+0309 was 
used as the phase calibrator and J0750+1231 as 
the bandpass and flux calibrator. A typical ALMA 
Band-3 calibration uncertainty of 5\% is adopted. 
The total on-source time was 118.4\,min, with an 
additional overhead time of 47.7\,min, 
resulting in a sensitivity of $\sim$\,0.12\,mJy/beam 
in 50\,km\,s$^{-1}$ velocity bins.

\begin{figure}[htbp]
\centering
\includegraphics[scale=0.79]{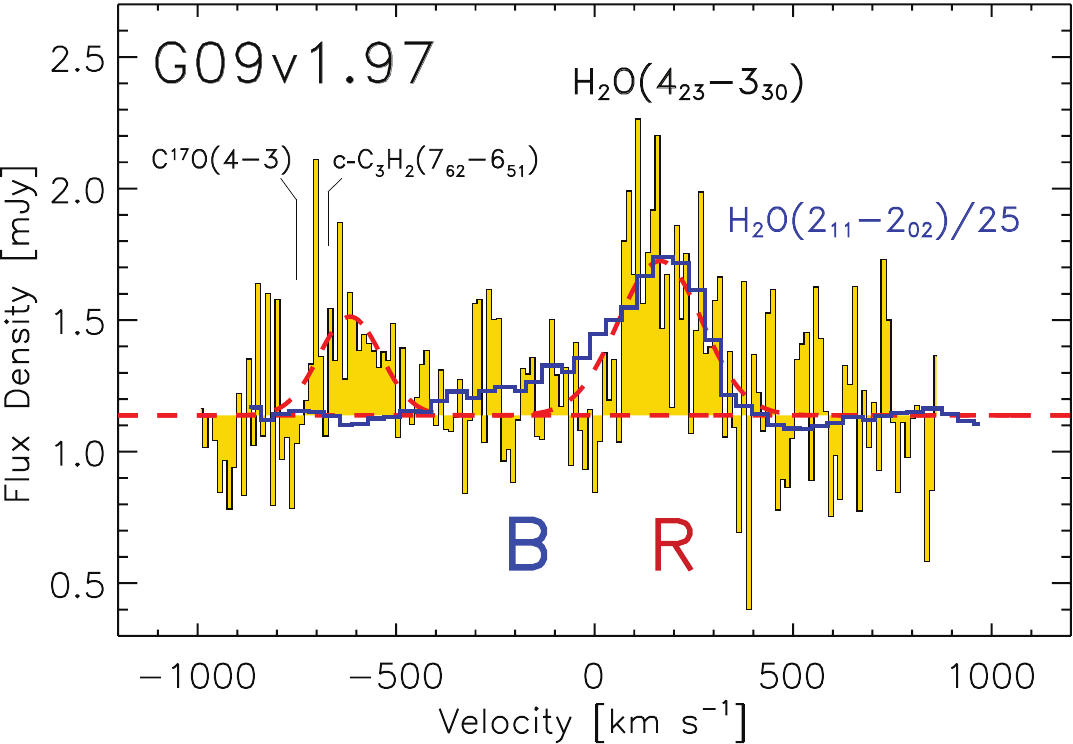}
\vspace{-0.25cm}
\caption
{
Spatially integrated spectrum of the 
448\,GHz \hto\ line of G09v1.97 (yellow histograms). 
B and R correspond to the blue- and red-shifted 
components of G09v1.97. The overlaid blue 
line shows the observed \htot211202\ line 
\citepalias{2019A&A...624A.138Y} after 
scaling down its flux density by a factor 
of 25. The dashed red lines show Gaussian 
fitting to the emission lines. 
Note that close to the 448\,GHz \hto\ line, 
we also tentatively detect a 2.3-$\sigma$ 
emission line (at the velocity 
of the R component) at $\sim$\,449\,GHz, 
which may be either C$^{17}$O(4--3) or 
c-C$_3$H$_2$(\t762651). If the emission 
is indeed C$^{17}$O(4--3), the integrated 
flux density ratio of C$^{16}$O(4--3)/C$^{17}$O(4--3) 
would be $\sim$\,110, which will be discussed 
in Yang\,et\,al.\,(in prep.). 
The dust continuum at rest-frame 615\,$\mu$m 
is also detected with a flux density of 
$1.13\pm0.04$\,mJy.
}
 \label{fig:h2o-spec}
 \end{figure}

The data were calibrated using the ALMA calibration 
pipelines with minor flagging. The calibrated 
data were further processed with imaging and 
\texttt{CLEAN}ing using \texttt{tclean} procedure 
of the \texttt{CASA} software \citep{2007ASPC..376..127M} 
version 5.4.0, with a natural weighting (synthesis 
beamsize of 2.46\arcsec$\times$2.03\arcsec and 
$PA$\,=\,78.2$^{\circ}$) to maximize the signal 
to noise ratio. The beamsize is unable to resolve 
the source, which has the largest angular structure 
of $\sim$\,2\arcsec \citepalias{2019A&A...624A.138Y}.
The spectrum was then extracted from the spatially 
integrated emission over the entire source 
(Fig.\,\ref{fig:h2o-spec}).

\setlength{\tabcolsep}{0.82em}
\begin{table*}[!htbp]
\centering 
\small 
\caption{Spatially integrated \hto\ line properties of G09v1.97-R.}
\begin{tabular}{rcrccccc}
\toprule
 Line\;\;\;\;\;\;& $E_\mathrm{u}$& $\nu_\mathrm{rest}$\;\;\;& $\mu_{\mathrm{R-line}} I_\mathrm{H_{2}O}$ & $\Delta{V}_\mathrm{line}$ & $\mu_{\mathrm{R-line}} L_\mathrm{Line}$\;  & $\mu_{\mathrm{R-line}} L'_\mathrm{Line}$    &   reference   \\
            &      (K)      &      (GHz)\;             & (Jy\,\kms)        & (\kms)                    &  (10$^{8}$\,\lsun)       &    (10$^{10}$\,\kkmspc)   &               \\
\midrule                                                                                              
para-\htot211202 &      137      &     752.033              & $4.2\pm0.6$\rlap{$^a$} &   $257\pm27$\rlap{$^b$}   &     $7.6\pm0.4$          &       $5.6\pm0.3$         &   Y16, Y19    \\
ortho-\htot321312 &      305      &    1162.912              & $3.7\pm0.4$       &   $234\pm34$              &     $10.3\pm1.1$\;       &       $2.1\pm0.2$         &   Y16         \\
para-\htot423330 &      433      &     448.001              & $0.16\pm0.03$     &   $250$\rlap{$^c$}        &  \;$0.17\pm0.03$         &       $0.60\pm0.11$       &   This work   \\
ortho-\htot422413 &      454      &    1207.639              & $1.8\pm0.6$       &   $328\pm139$             &      $5.2\pm1.7$         &       $0.9\pm0.3$         &   Y17         \\ 
\bottomrule
\end{tabular}
\tablefoot{  
Observed \hto\ line properties of G09v1.97-R. 
We adopt $\mu_{\mathrm{R-line}}$\,=\,20 for 
the magnification of the velocity integrated 
line fluxes of G09v1.97-R \citepalias{2019A&A...624A.138Y}.  
$^{(a)}$: Here, we include an additional 15\% 
uncertainty considering the possible 
contamination from the interacting region 
\citepalias{2019A&A...624A.138Y}.
$^{(b)}$: The value is taken from the NOEMA 
detection by \cite{2016A&A...595A..80Y} of  
the red-shifted component, for 
the consistency in comparison with the other 
lines detected here, which are also dominated 
by the red-shifted component (G09v1.97-R).
$^{(c)}$: Here, we adopt a fixed linewidth 
for G09v1.97-R of 250\,\kms, according to 
\cite{2016A&A...595A..80Y}.
References: Y16=\cite{2016A&A...595A..80Y};  
Y17=\cite{2017PhDT........21Y}; 
Y19=\cite{2019A&A...624A.138Y}.
}
\label{table:src-spec}
\end{table*}
\normalsize

\section{Analysis and Discussion}
\label{section:results}

As shown by the high-angular resolution 
observations in \citetalias{2019A&A...624A.138Y}, both 
the CO(6--5) and \htot211202\ lines of G09v1.97 consist 
mainly of B (blue-shifted) and R (red-shifted) components 
(Fig.\,\ref{fig:h2o-spec}), with linewidths of 
$\sim$\,300\,km\,s$^{-1}$. The B component originates 
exclusively from the approaching galaxy G09v1.97-B, 
while the R component arises from the receding galaxy 
G97v1.97-R. Therefore, the contribution to the 
spectrum from each merger companion can be 
disentangled without spatially resolved observations. 
The lensing magnification for the line varies
from $\sim$\,5 to $\sim$\,22 as a function of velocity,  
from the blue-shifted to the red-shifted velocity components, 
due to a velocity gradient from south to north 
\citepalias{2019A&A...624A.138Y}. As a result, R 
is a factor of $\gtrsim$\,4 brighter than B in the spectrum, 
causing an extremely asymmetric line profile 
(blue histogram in Fig.\,\ref{fig:h2o-spec}). 
The 448\,GHz \hto\ line is detected with $\gtrsim$\,5-$\sigma$ 
significance, but only in the red-shifted 
component (Fig.\,\ref{fig:h2o-spec}), namely only 
in G09v1.97-R. Assuming a similar flux ratio of 
\htot423330/\htot211202\ in R and B, the 448\,GHz \hto\ 
line in G09v1.97-B is thus buried below the noise level. 
Therefore, we associate the detected 448\,GHz \hto\ line 
only with G09v1.97-R. We have then corrected the lensing 
magnification for the \hto\ line fluxes of G09v1.97-R, as 
listed in Table\,\ref{table:src-spec}. The table also 
includes the observational results on the \htot321312\ 
\citep{2016A&A...595A..80Y} and \htot422413\ lines  
\citep{2017PhDT........21Y} for G09v1.97-R, which were 
obtained with NOEMA with similar spatial resolution of 
$\gtrsim$\,2\arcsec (unresolved).
The continuum measurements, which are used below 
to model the \hto\ emission, have also been corrected to 
account only for G09v1.97-R. The limited spatial resolution 
of the current highest spatial resolution dust continuum 
images \citepalias{2019A&A...624A.138Y} does not allow us to 
disentangle the fluxes from the merger companions. Nevertheless, 
the relative contribution by G09v1.97-R can be approximated 
by using the tight correlation between \lco65\ and \lir\ 
\citep{2015ApJ...810L..14L, 2017A&A...608A.144Y}.
Since the intrinsic line fluxes from R are about 50\% of the 
total, and the magnifications R and B are 
$\mu_{\mathrm{R}}$\,$\approx$\,20 and $\mu_{\mathrm{B}}$\,$\approx$\,5, 
respectively \citepalias{2019A&A...624A.138Y},
the intrinsic continuum flux densities from G09v1.97-R
are a factor $\approx25$ smaller than the observed values
integrated over the entire G09v1.97 system. Therefore, 
here we scale down the measured continuum fluxes 
by a factor of 25 and include an uncertainty of 20\% 
associated with the flux scaling.

\begin{figure}[htbp]
\centering
\includegraphics[scale=0.57]{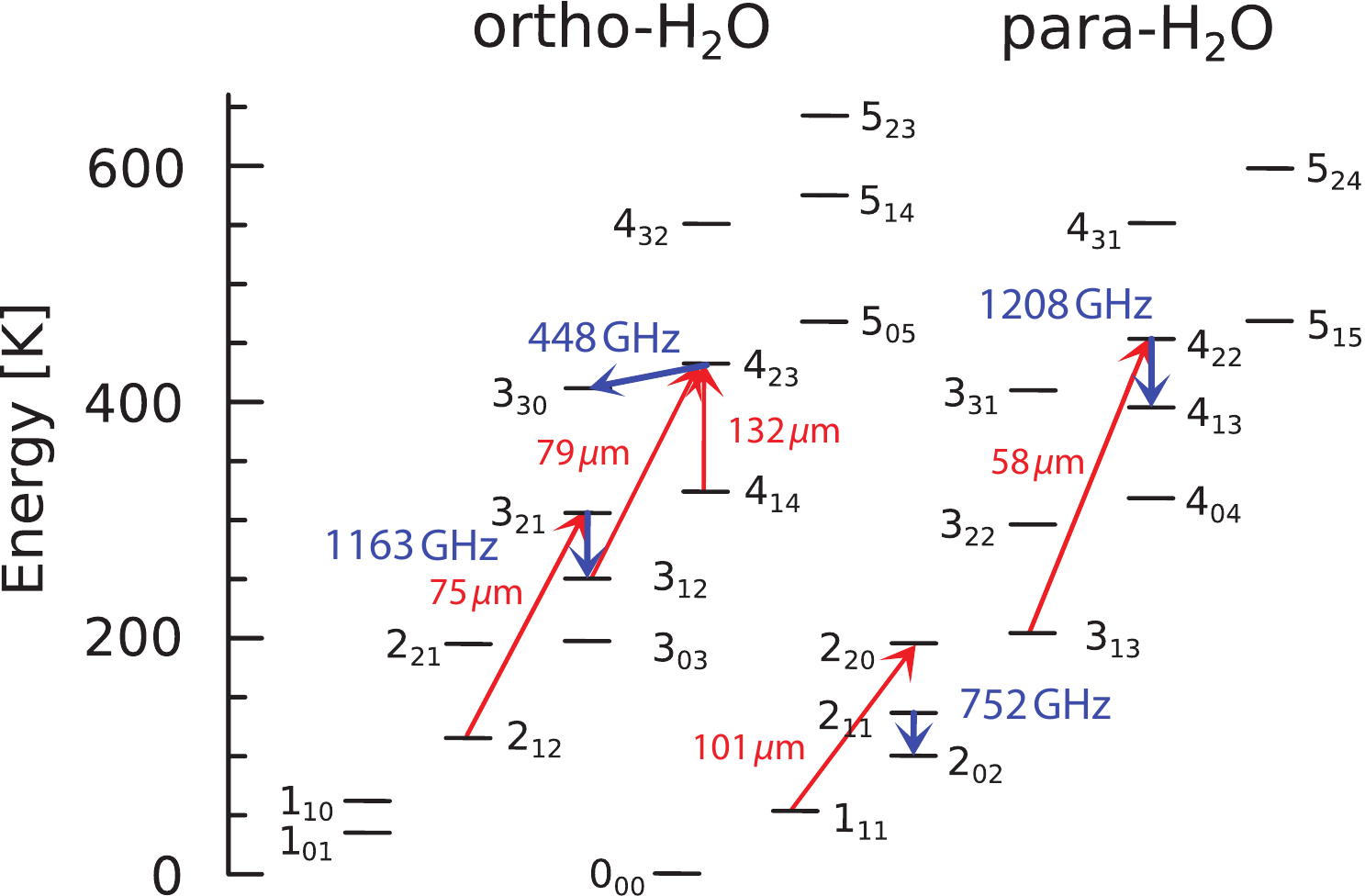}
\vspace{-0.2cm}
\caption
{
Energy level diagram of the main para- and ortho-\hto\ 
transitions. Blue arrows indicate the emission lines 
modeled in this work. Red arrows show the 
transitions responsible for the associated 
radiative pumping. The frequencies of the lines 
and the wavelengths of the far-IR photons 
are labeled.
}
\label{fig:h2o-level}
\end{figure}

In G09v1.97-R, the 448\,GHz ortho-\htot423330\ line has 
a luminosity of $8.5 \times 10^{5}$\,\lsun, which is a 
fraction of 10$^{-7}$ of the total IR luminosity. 
Besides the 448\,GHz \hto\ line, three additional 
lines of \hto\ are previously detected in G09v1.97-R
(Fig.\,\ref{fig:h2o-level} and Table\,\ref{table:src-spec}):  
the ortho-\htot321312, the para-\htot211202, and the 
para-\htot422413\ transitions. Interestingly, the \t423330\ 
and \t422413\ transitions have $J_{\mathrm{upper}}$\,=\,4 
levels with very similar energies (425\,K versus 433\,K),  
but have $A$-Einstein coefficients for spontaneous emission 
that differ by a factor of $\approx$\,500\,:\,$5.4\times10^{-5}$ and
$2.8\times10^{-2}$\,s$^{-1}$, respectively \citep{1998JQSRT..60..883P}. 
Assuming that the upper-level populations of both 
lines do not strongly differ from the ortho-to-para ratio of 
3 appropriate for warm regions, their flux ratio  
(in Jy\,km\,s$^{-1}$) in optically thin conditions should be  
$F(1208\,\mathrm{GHz})/F(448\,\mathrm{GHz})\sim A_{ul}(1208\,\mathrm{GHz})/(3A_{ul}(448\,\mathrm{GHz}))\sim170$.
The observed flux ratio in G09v1.97-R is only $\approx$\,11 
(Table\,\ref{table:src-spec}), similar to the value of $\approx$\,15  
found in ESO\,320-G030 (Gonz\'alez-Alfonso\,et\,al., in prep.), 
indicating that the ortho-\htot422413\ line is strongly saturated 
in both sources. Similar to the situation found in buried 
nuclei of local (ultra)-luminous IR galaxies  
((U)LIRGs), high radiation and column densities in a nuclear
component are required to pump the $J_{\mathrm{upper}}$\,=\,4 
levels and to account for the corresponding strong \hto\ 
emission from G09v1.97-R.

To estimate the physical conditions and structure of 
the ISM in G09v1.97-R from the observed \hto\ and 
dust continuum emission, a library of model components has 
been developed following the method described in 
\cite{2014A&A...567A..91G}. We assume for each component 
a spherically symmetric source with uniform physical 
properties: the dust temperature $T_{\mathrm{dust}}$, 
the continuum optical depth at 100\,$\mu$m along a radial 
path $\tau_{100}$, the column density of \hto\ along 
a radial path $N_{\mathrm{H_{2}O}}$, the H$_2$ density 
$n_{\mathrm{H_{2}}}$, the velocity dispersion $\Delta V$, 
and the gas temperature $T_{\mathrm{gas}}$. The physical 
parameters modified from model to model are 
$T_{\mathrm{dust}}$, $\tau_{100}$, $N_{\mathrm{H_{2}O}}$, 
$n_{\mathrm{H_{2}}}$, and we keep fixed 
$\Delta V$\,=\,100\,km\,s$^{-1}$ and $T_{\mathrm{gas}}$\,=\,150\,K 
(consistent with the results from CO excitation, 
\citealt{2017A&A...608A.144Y}).
The model components are classified into groups according 
to their physical parameters, each group covering a regular 
grid in the parameter space ($T_{\mathrm{dust}}$, $\tau_{100}$, 
$N_{\mathrm{H_{2}O}}$, $n_{\mathrm{H_{2}}}$). Once the 
model components are created, all available \hto\ line fluxes 
and continuum flux densities are fitted simultaneously, 
by using a number $N_\mathrm{C}$ of model components (up to 2 components 
per group) and checking all possible combinations among them. 
Because the intrinsic line and continuum fluxes (in Jy\,km\,s$^{-1}$ 
and Jy, respectively) scale as $(1+z)\,(R/D_L)^{2}$, where $R$ 
is the source radius and $D_L=32.7$\,Gpc is the luminosity distance, 
a $\chi^2$ minimization procedure is used to determine the source 
radius $R$ of each component for all combinations of model components. 
Our best model fit corresponds to the combination that yields the lowest
$\chi^2$, while the results for all combinations enable us 
to calculate the likelihood distribution of the free physical 
parameters \citep[e.g.,][]{2003ApJ...587..171W}, i.e., 
$T_{\mathrm{dust}}$, $\tau_{100}$, $N_{\mathrm{H_{2}O}}$, 
$n_{\mathrm{H_{2}}}$, for each component. The derived parameters 
($R$, the \hto\ abundance relative to H nuclei $X_{\mathrm{H_{2}O}}$, and 
$L_{\mathrm{IR}}$) can also be inferred. More details will be 
given in Gonz\'alez-Alfonso\,et al. (in prep.).

\begin{figure*}[htbp]
\centering
\includegraphics[scale=0.575]{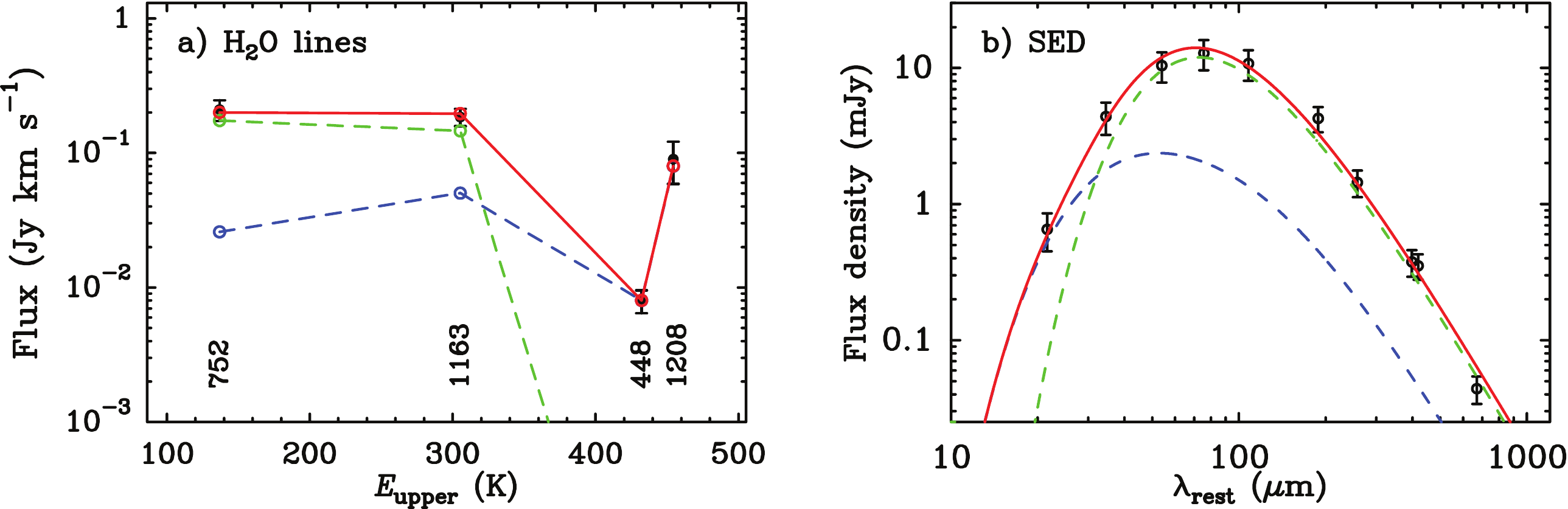}
\vspace{-0.25cm}
\caption
{
The best-fit two-component model of G09v1.97-R for 
a) the \hto\ lines and b) the dust continuum 
 (colored symbols and lines) compared with 
observations (black symbols with errorbars).
Predictions for the two individual ISM components 
are displayed with dashed green (the disk component) 
and blue lines (the core component); 
red is total. The physical parameters adopted from this model are 
indicated with arrows in Fig.\,\ref{fig:model-parameter}.
In panel a, the numbers indicate the rest frequencies 
of the lines in GHz. The observed \hto\ fluxes and continuum flux
densities have been corrected for magnification. The 
continuum fluxes have also 
been corrected to only account for the contribution by  
G09v1.97-R, as described in Section~\ref{section:obs_redu}
(see also \citetalias{2019A&A...624A.138Y}).
}
 \label{fig:model}
 \end{figure*}

\begin{figure*}[htbp]
\centering
\includegraphics[scale=0.5]{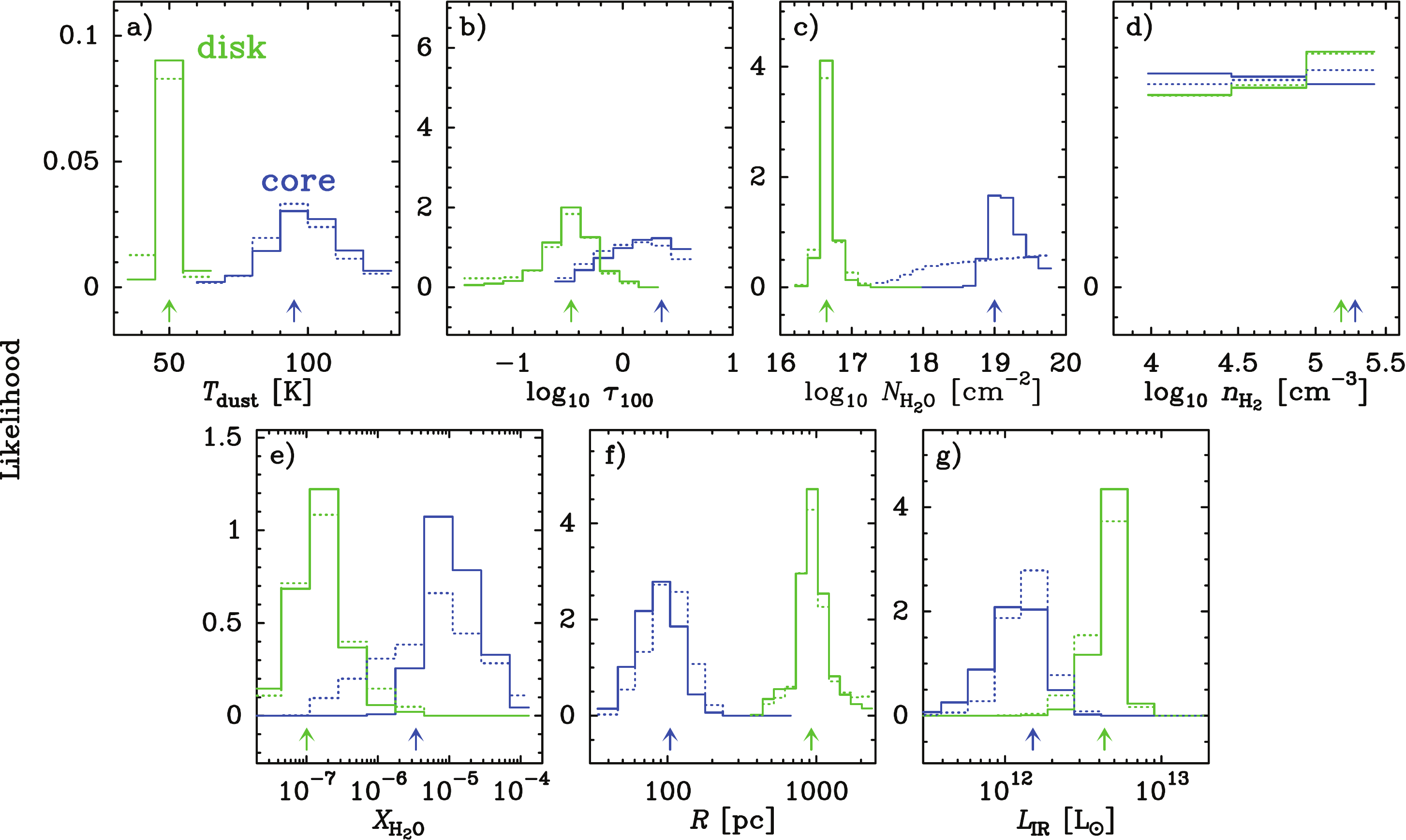}
\vspace{-0.25cm}
\caption{
The solid histograms indicate the likelihood distributions 
for the physical parameters of the two-component model  
(green and blue for the disk and core component, 
respectively). a) to d) are fitted parameters while e) 
to f) are derived parameters. 
Dotted histograms show the likelihood distributions 
of a fit that ignores the 448\,GHz \hto\ line, showing 
consistent results but a high uncertainty in 
$N_\mathrm{H_{2}O}$ and $X_\mathrm{H_{2}O}$. 
The arrows indicate the values of the best model fit, 
with results for the \hto\ fluxes and
continuum flux densities displayed in Fig.\,\ref{fig:model}.
}
\label{fig:model-parameter}
\end{figure*}

We first attempted to fit the \hto\ and continuum emission 
with a single-model component ($N_\mathrm{C}$\,=\,1), but 
results were unreliable with a best reduced chi-square 
value $\chi_\mathrm{red}^2$\,$\approx$\,4. This was indeed 
expected because the low-lying \hto\ $J_{\mathrm{upper}}$\,=\,2--3 
lines are expected to arise in more extended regions than 
the $J_{\mathrm{upper}}$\,=\,4 lines that trace 
buried regions \citep[e.g.,][]{2014A&A...567A..91G,2017A&A...601L...3P}. 
A better fit was found with $N_\mathrm{C}$\,=\,2 
components (Fig.\,\ref{fig:model}), with 
$\chi_\mathrm{red}^2$\,$\approx$\,0.9. In Fig.\,\ref{fig:model-parameter}, 
the arrows indicate the best-fit values, and 
solid histograms show their likelihood distributions. 
We find that the $J_{\mathrm{upper}}=4$ lines 
are formed in a very warm nuclear region (core) 
with size $R$\,$\sim$\,$100$\,pc and 
$T_{\mathrm{dust}}$\,$\sim$\,$100$\,K, which is most probably 
optically thick at far-IR wavelengths ($\tau_{100}$\,$\gtrsim$\,$1$). 
This nuclear core has a luminosity of 
$L_{\mathrm{IR}}$\,$\sim$\,$10^{12}$\,$L_{\odot}$, 
resulting in an extreme IR luminosity surface density 
$\Sigma_\mathrm{IR}$\,=\,$8\times10^{12}$\,\lsun\,kpc$^{-2}$. 
This translates to a surface star formation rate 
of $\sim$\,$1.1\times10^{3}$\,\msun\,yr$^{-1}$\,kpc$^{-2}$, 
if the contribution to $L_{\mathrm{IR}}$ by a possible 
obscured AGN is negligible, 
since no strong evidence of the presence of an AGN
has been found \citep{2016A&A...595A..80Y}. 
A large column density of water, 
$N_{\mathrm{H_{2}O}}$\,$\sim$\,$10^{19}$\,cm$^{-2}$ 
($X_\mathrm{H_{2}O}$\,$\sim$\,$10^{-5}$), is found 
in this core component, which is more than one order 
of magnitude higher than in the \textit{z}\,=\,3.9 
quasar host galaxy APM\,08279+5255 \citep{2011ApJ...741L..38V}. 
In addition, the $J_{\mathrm{upper}}$\,=\,2--3 lines, pumped 
by absorption of dust-emitted 100 and 75\,$\mu$m photons 
(Fig.\,\ref{fig:h2o-level}), require an optically thin 
($\tau_{100}$\,$\sim$\,0.3) and more 
extended region of radius $\sim1$\,kpc (the disk component), 
remarkably comparable to the projected half-light effective 
radius traced by the \co65\ and \htot211202\ line 
emission (0.8\,kpc, \citetalias{2019A&A...624A.138Y}), 
and also to the size of the averaged 870\,$\mu$m dust continuum of 
SMGs \citep[$\sim$\,1\,kpc,][]{2019MNRAS.490.4956G}.
With $T_{\mathrm{dust}}$\,$\sim$\,$50$\,K, the disk   
component dominates the luminosity with 
$L_{\mathrm{IR}}$\,$\sim$\,$5\times10^{12}$\,$L_{\odot}$ 
and $\Sigma_\mathrm{IR}=4\times10^{11}$\,\lsun\,kpc$^{-2}$ 
(corresponding to $\sim$\,60\,\msun\,yr$^{-1}$\,kpc$^{-2}$), 
and and has lower column density and abundance of water than 
the core, with $N_{\mathrm{H_{2}O}}$\,$\sim$\,$10^{16}$\,cm$^{-2}$ 
($X_\mathrm{H_{2}O}$\,$\sim$\,$10^{-7}$).

While the effect of the cosmic microwave background (CMB) 
on the dust SED is included in Fig.\,\ref{fig:model}b 
following \cite{2013ApJ...766...13D} (see also
\citealt{2016RSOS....360025Z}), the correction is small 
as $T_{\mathrm{dust}}$\,$>>$\,$T_{\mathrm{CMB}}$\,=\,$12.5$\,K. 
The excitation of \hto\ by the CMB, however, is not included 
in our models, but we have checked that, for our best fit 
model (Fig.\,\ref{fig:model}a), it has negligible effect 
on the excitation and flux of the observed lines.
A more important source of uncertainty is that our models
do not include the effects of spatially varying $T_{\mathrm{dust}}$ 
that are expected in regions that are optically thick in the 
far-IR \citep{2019ApJ...882..153G}. 
An increasing $T_{\mathrm{dust}}$ outside-in would facilitate the
\hto\ excitation within the core component, and would thus decrease
to some extent the inferred $N_{\mathrm{H_{2}O}}$.
This effect becomes most relevant for extremely buried
sources with $N_{\mathrm{H_{2}}}$ approaching $10^{25}$\,cm$^{-2}$
($\tau_{100}\gtrsim10$), which is not excluded for the nuclear 
core of G09v1.97-R (Fig.\,\ref{fig:model-parameter}b).

What is the role of the 448\,GHz \hto\ line in 
constraining the model? The dotted histograms 
in Fig.\,\ref{fig:model-parameter} show that, if the 448\,GHz 
\hto\ line is excluded, the likelihood distributions 
of the physical parameters remain similar, but show a less 
informative distribution for $N_{\mathrm{H_{2}O}}$ 
(and $X_\mathrm{H_{2}O}$). In this case,
the parameters of the nucleus are determined by the 
1208\,GHz \hto\ line, which is optically thick 
($\tau_{1208-\mathrm{H{_2}O}}$\,=\,$27.5$). 
The detection of the more optically thin 448\,GHz \hto\ 
line ($\tau_{448-\mathrm{H{_2}O}}$\,=\,$1.0$) 
confirms the occurrence of the warm, nuclear region, 
and enables a significantly more accurate estimate 
of its $N_{\mathrm{H_{2}O}}$ (and $X_\mathrm{H_{2}O}$).  
On the other hand, the flat distributions of $n_{\mathrm{H_{2}}}$ 
in both regions indicate that the excitation of these 
\hto\ lines is insensitive to $n_{\mathrm{H_{2}}}$.

Our models, which fit the \hto\ line and dust continuum 
fluxes simultaneously, are sensitive to $N_{\mathrm{H_{2}O}}$ 
and $\tau_{100}$. These two parameters, treated as 
independent, are linked as
$X_{\mathrm{H_{2}O}}$\,=\,$N_{\mathrm{H_{2}O}}/(1.3\times10^{24} \tau_{100} (\mathit{GDR}/100))$, where $\mathit{GDR}$ 
is the gas-to-dust ratio by mass.
For the core component, our best fit yields 
$N_{\mathrm{H_{2}O}}$/$\tau_{100}$\,$\sim$\,$5\times10^{18}$\,cm$^{-2}$.
With the assumption of a $\mathit{GDR}$\,=\,100, we obtain
$X_{\mathrm{H_{2}O}}$\,$\sim$\,$(0.3\text{--}5)\times10^{-5}$ 
(90\% confidence interval, Fig.\,\ref{fig:model-parameter}e), 
similar to the abundance in buried galactic nuclei 
of local (U)LIRGs \citep{2012A&A...541A...4G}.
The similarity of the $N_{\mathrm{H_{2}O}}/\tau_{100}$ ratio
in the nuclear cores of SMGs and of local (U)LIRGs appears 
to indicate a fundamental similarity of the gas metallicity 
to dust ratio in buried galactic nuclei across cosmic times.

With the \hto\ excitation models, we are able to ``resolve'' 
two distinct components with significantly different properties 
(e.g., $T_\mathrm{dust}$ and $\tau_{100}$). 
These differences reveal the complexity of the ISM structure  
in SMGs and suggest caution when deriving the physical properties 
from spatially unresolved observations. 
For example, an increase of dust 
temperature and optical depth towards the nuclear region 
can significantly alter the calculation of size and mass of the 
ISM dust \citep[e.g.,][]{2017A&A...602A..54M,2018ApJ...863...56C}. 
While the disk component, dominating the total IR 
emission, has a value of the star formation rate surface 
density typical of SMGs \citep{2013ApJ...768...91H}, the nuclear core 
shows more extreme conditions. The properties of the 
optically thick core show excellent agreement with the 
characteristic \lir\ surface density ($\sim$\,$10^{13}$\,\lsun\,kpc$^{-2}$) 
and dust temperature ($\sim$\,90\,K) of Eddington-limited
starburst models \citep{2005ApJ...630..167T}, indicating the 
importance of radiation pressure on dust in regulating the ISM in the 
inner 100\,pc region of G09v1.97-R. Although we cannot rule 
out a contribution to the far-IR from an obscured AGN, in 
addition to the arguments disfavoring a dominant AGN power source 
discussed in the introduction and in \citetalias{2019A&A...624A.138Y}, 
we also find no strong evidence for the presence of a powerful  
AGN in G09v1.97-R by checking the $q$ parameter \citep{1992ARA&A..30..575C} 
and mid-IR excess \citep[see discussions in][]{2016A&A...595A..80Y} 
thus strengthening the case for Eddington-limited starbursts.
Nevertheless, further observations are needed to quantify  
the contribution from an AGN.

\section{Conclusions}
\label{section:conclusion}

We report the first detection of the 448\,GHz ortho-\htot423330\ 
line at high redshift, in a {\it z}\,=\,3.63 lensed submillimeter 
galaxy, G09v1.97-R. In combination with three other transitions 
of \hto\ and dust emission, we have built a radiative transfer 
model for the \hto\ excitation and dust emission. The 
model decomposes the dust continuum of the SMG into two components, 
a $\sim$\,$1$\,kpc optically thin ($\tau_{100}$\,$\sim$\,0.3) 
disk component with $T_{\rm dust}$\,$=$\,$50$\,K emitting a total 
infrared luminosity of $\sim$\,$5\times10^{12}$\,\lsun, and 
a $\sim$\,$0.1$\,kpc heavily dust-obscured ($\tau_{100}$\,$\gtrsim$\,1) 
nuclear core with very warm dust (100\,K) and an infrared luminosity of 
$\sim10^{12}$\,\lsun. The water abundance in the core 
($X_\mathrm{H_{2}O}$\,$\sim$\,$5\times10^{-6}$) is more than one 
order of magnitude higher than in the more extended disk. 
The core possesses a surface star formation rate of 
$\Sigma_\mathit{SFR}$\,$=$\,$1.1\times10^{3}$\,\msun\,yr$^{-1}$\,kpc$^{-2}$, 
which is $\sim$\,20 times higher than that of the disk. The 
ISM properties of the nucleus of G09v1.97-R resemble the 
characteristic conditions of an Eddington-limited starburst, 
indicating that radiation pressure on dust plays an 
essential role in supporting the ISM. 

The optically thin 448\,GHz \hto\ line is a powerful tool 
for the study of SMGs around redshift 2--4 (with ALMA Bands 3 and 4). 
The multi-component structure derived from the \hto\ excitation 
model reveals the complex nature and morphology of SMGs and 
may offer clues about the evolutionary link of the SMGs to 
massive elliptical galaxies today, once high-resolution near-infrared 
observations of the stellar component are possible and become 
available with telescopes such as JWST.

\begin{acknowledgement}
We thank the anonymous referee for very helpful suggestions.
C.Y. acknowledges support from an ESO Fellowship.  
E.G.-A. is a Research Associate at the Harvard-Smithsonian 
Center for Astrophysics, and thanks the Spanish 
Ministerio de Econom\'{\i}a y Competitividad for 
support under project ESP2017-86582-C4-1-R. 
E.G.-A. also thanks the support from the ESO Chile 
Scientific Visitor Programme. 
M.P.S. acknowledges support from the Comunidad 
de Madrid through Atracci\'on de Talento 
Investigador Grant 2018-T1/TIC-11035. 
This paper makes use of the following ALMA 
data: ADS/JAO.ALMA\#2018.1.00797.S.
ALMA is a partnership of ESO (representing its member 
states), NSF (USA) and NINS (Japan), together with NRC 
(Canada), MOST and ASIAA (Taiwan), and KASI (Republic 
of Korea), in cooperation with the Republic of Chile. 
The Joint ALMA Observatory is operated by ESO, 
AUI/NRAO and NAOJ.
\end{acknowledgement}

\small{
\bibliographystyle{aa_url}
\bibliography{2019-g09-h2o} 
}

%
%

\end{CJK*}
\end{document}